\documentclass[pre]{revtex4-2}
\usepackage{amssymb}
\usepackage{amsmath}
\usepackage{graphicx}
\usepackage{hyperref}
\usepackage{bm}
\ifx\affiliation\undefined 
\def\affiliation#1{\date{\normalsize #1}}
\usepackage[margin=1in]{geometry}
\usepackage[sort&compress,numbers]{natbib}
\else 
\fi

\begin{document}
\title{How can slow plasma electron holes exist?}
\author{I H Hutchinson}
\affiliation{Plasma Science and Fusion Center, Massachusetts
  Institute of Technology, Cambridge, MA, USA} 

\ifx\altaffiliation\undefined\maketitle\fi 
\begin{abstract}
  One dimensional analysis is presented of solitary positive
  potential plasma structures whose velocity lies within the range of ion
  distribution velocities that are strongly populated: so called
  ``slow'' electron holes. It is shown that to avoid the
  self-acceleration of the hole velocity away from ion velocities it
  must lie within a local minimum in the ion velocity distribution.
  Quantitative criteria for the existence of stable equilibria are
  obtained. The background ion distributions required are generally
  stable to ion-ion modes unless the electron temperature is much
  higher than the ion temperature. Since slow positive potential
  solitons are shown not to be possible without a significant
  contribution from trapped electrons, it seems highly likely that
  such observed slow potential structures are indeed electron holes.
\end{abstract}
\ifx\altaffiliation\undefined\else\maketitle\fi  

\section{Introduction}

Solitary positive potential structures are observed by satellites in
some space plasmas to have speeds comparable to the typical ion
thermal speed, even lying within the strongly populated velocities of
the ion distribution\cite{Graham2016,Steinvall2019,Lotekar2020}; that
is what is meant here by calling the structures ``slow''. A candidate
explanation of these structures is that they are ``slow electron
holes'', in which the positive potential is sustained by a deficit of
trapped electrons. However, till now it has been unclear theoretically
whether, or under what circumstances, slow electron holes can
exist. The purpose of the present study is to discover the theoretical
conditions for the existence of plasma-sustained steady slow solitary
positive potential structures, including electron holes, and identify
the mechanisms that control them.

\begin{figure}[htp]
  \centering
  \includegraphics[width=0.5\hsize]{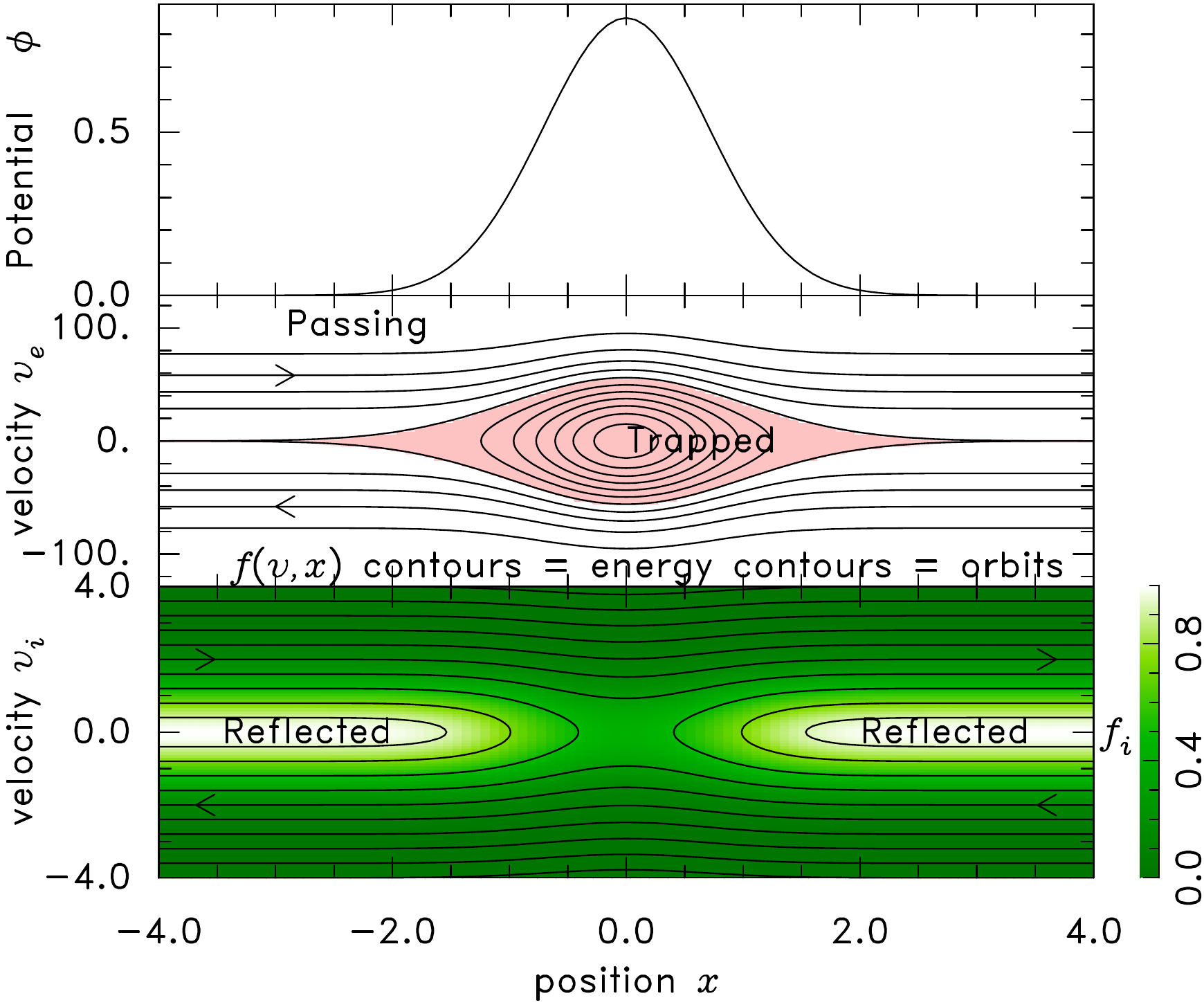}
  \caption{Schematic of a slow electron hole and the corresponding
    electron and ion phase-space density contours.}
  \label{fig:electronholeorbits}
\end{figure}
Figure \ref{fig:electronholeorbits} shows contours of electron and
velocity distribution functions $f(v)$, for a presumed steady slow
potential peak, in their respective $x,v$ phase-spaces (using
conveniently normalized units). For \emph{trapped} electron orbits,
$f_e(v)$ is determined by conditions during the structure's formation
and has lower value than the nearby passing orbits which are
determined by the boundary conditions. This results in a more negative
central electron density than ion density and causes the potential
peak. (Electron contour values are not important to this illustrative
discussion and not shown.) The ion distribution is everywhere
determined by the distant distribution function and the fact (arising
from Vlasov's equation) that $f$ is constant along orbits. Orbits have
constant energy for a steady potential. Ions are reflected by
the hole if their speed in the hole frame is small enough. The
illustrative case shown corresponds to Maxwellian ion distribution at
large $|x|$, with zero average velocity in the hole frame. That is,
this hole has zero velocity in the ion frame.

The reasons to question whether slow electron holes can exist are to
do with the interaction of their positive potential peak with the
ions. Classic electron holes move at speeds, relative to ions, up to
of order the electron thermal speed
$v_{te}$\cite{Schamel1986a,Hutchinson2017}. And when they are at more
than a very small fraction of $v_{te}$, the ion perturbation is small
because the duration of any moving electron hole's interaction with an
ion is much smaller than the typical response time of the (far
heavier) ions. Ion response can then often be completely ignored. As
has been extensively discussed, for example in the original paper on
BGK-modes\cite{Bernstein1957} and the electron hole review
literature\cite{Hutchinson2017} the detailed shape of the hole has
considerable latitude to adjust itself to the details of the trapped
electron velocity distribution, and is only a minor consideration here,
limiting the discussion to potentials with a single maximum.

For slower holes ion interaction gradually becomes important.  When
hole speed is less than a few (up to about $(m_i/m_e)^{1/4}$) times
the ion acoustic speed ($c_s=\sqrt{T_e/m_i}$), the ion interaction is
significant, and the electron hole speed resists approaching the ion
speed\cite{Hutchinson2016,Zhou2016}, maintaining a velocity
(difference) greater than a minimum that increases with hole
potential. If the hole speed is less than that minimum, an oscillatory
instability in the hole speed arises\cite{Zhou2017} and there is
therefore a forbidden region of hole speed. This forbidden velocity
region has a lower limit that is at approximately the ion-acoustic
soliton speed, which is just above $c_s$ (depending on peak
potential\cite{Davidson1972}). At that specific speed, an entity
usually called a coupled hole-soliton (CHS) is known from
simulations\cite{Saeki1991,Saeki1998,Zhou2018} to exist. In effect the
electron hole is trapped in, and enhances, the positive potential
produced at that speed by the positive ion density perturbation of the
ion-acoustic soliton. A CHS generally moves faster than the ion
thermal speed, provided the electron temperature is greater than ion
temperature, so there are few ions in the distribution at the CHS
speed, and Landau damping can be small.

It is emphasized that none these known types of theoretical holes
qualifies for the present meaning of ``slow''. Neither do holes
produced by the Bunemann instability when there is substantial drift
between ion and electron populations, invoked by Norgren et
al\cite{Norgren2015,Norgren2015a} to explain their space
observations. Holes produced\cite{Drake2003,Khotyaintsev2010,Zhou2018}
by Bunemann instability usually have speeds (relative to ions)
$\ll v_{te}$ but not $<c_s\simeq\sqrt{m_e/m_i}\,v_{te}$, let alone
$\sim v_{ti}$.  Instead, the present paper addresses the ``Group 3''
electron holes observed by Steinvall et al\cite{Steinvall2019} (``on
the magnetospheric side of the magnetopause'') that have speeds
relative to ions below $c_s$ (Group 1 speeds exceed the oscillatory
instability threshold, and Group 2 are consistent with being CHS
type). A fraction of the observations of Graham et al\cite{Graham2016}
(``near the magnetopause'') and of the blue points in Figure 4 of
Lotekar et al\cite{Lotekar2020} (magnetotail) also are slow in the
present sense.

Simulations that initialize an electron hole at speeds of order the
ion thermal speed $v_{ti}$ or less in an initially uniform ion
background, observe a remarkable and rapid ``self-acceleration'' of
the
hole\cite{Saeki1991,Muschietti1999,Eliasson2004,Eliasson2006,Zhou2016}. The
growing negative ion density (and hence charge) perturbation caused by
the repulsion of ions from the positive potential of the hole repells
the electron hole, because an electron hole's dynamics as a composite
entity are such that it has an effective charge to mass ratio equal to
that of the electron\cite{Haakonsen2015,Hutchinson2016}. A short time
after initialization, it moves away at speeds much larger than
$v_{ti}$. Thus, past simulation attempts have failed to produce steady
slow electron holes.

The novelty and complexity of the present analysis in comparison with
the prior treatments of ion-acoustic solitons and electron holes is
that it \emph{requires a kinetic} (rather than fluid, e.g.\cite{Kakad2016})
treatment of the ions in equilibrium. Concerning past kinetic electron
hole analysis (e.g.\cite{Dupree1982}) and simulations
(e.g.\cite{Eliasson2004}), the key difference is that the present
analysis shows that for \emph{slow} positive structures to exist
stably, the background ion velocity distribution generally cannot be
``single-humped''. It must instead possess at least two maxima.
Indeed, it is shown that slow positive solitary potentials sustained
by trapped electron deficit (1) cannot persist in single-humped ion
distributions; (2) can persist only when the velocity of the electron
hole lies within a local minimum of the ion distribution function; but (3)
do not require background distributions that are ion-ion unstable,
provided the electron temperature is not very high. All the discussion
here is one-dimensional, and multidimensional stability is beyond the
present scope. The conclusion therefore is that, from a
one-dimensional perspective, slow electron holes can exist, requiring
distinctively non-thermal external ion distributions; but those
distributions are not themselves unstable and can therefore persist
for substantial time durations.

These theoretical characteristics are valuable for identifying the
nature of slow solitary potential peaks observed in plasmas, and for
indicating the presence of double-humped ion distributions. Recent
analysis reported elsewhere [Kamaletdinov et al 2021, submitted to
Physical Review Letters] of satelite measurements confirm the
characteristics for slow electron holes observed in the plasma sheet
boundary layer.

Section \ref{sec1} addresses ion distribution functions that have
reflectional symmetry in some reference frame. The simplifications of
symmetry make it easier to understand the concepts introduced and
permit straightforward proofs concerning stability and
equilibrium. Section \ref{sec2} generalizes these results to
asymmetric ion distributions, and section \ref{sec3} addresses the
question of the linear stability of the uniform background ion
distributions found to be necessary for the existence of slow electron
holes.

\section{Symmetric Distribution Functions}\label{sec1}

Consider a steady solitary positive potential structure in one
dimension: $\phi(x)$, possessing a single maximum $\phi=\psi$ at
position $x=0$, and tending to the same potential $\phi=\phi_\infty=0$
at distant positions $x\to \pm \infty$. For motion in this single
dimension, suppose the distribution function of ions approaching the
potential structure from the distant plasma to be given as
$f_\infty(v_\infty)$, in a frame of reference in which the structure
is stationary.

 A collisionless ion equilibrium
satisfies the steady Vlasov equation, giving conservation of
distribution function and energy on orbits, leading to
\begin{equation}
  f(x,v)=f_\infty(\infty,v_\infty) \qquad\mbox{where}\qquad 
  v^2/2+\phi(x) = v_\infty^2/2+\phi_\infty. 
\end{equation}
In this paper, to abbreviate the equations we mostly work in
conveniently scaled units: energy normalized to thermal energy for a
reference temperature $T_0$, length normalized to Debye length
$\lambda_D=\sqrt{\epsilon_0 T_0/e^2n}$, and velocity to ion thermal
speed $\sqrt{T_0/m_i}$. In these units, the ion mass and charge are
unity. Where numerical values of quantities like distribution
function, density, or force-density etc., are presented, they are for
unit background density $n_\infty$.

When the ion distribution function \emph{in the rest frame of the
  structure} is not symmetric, very substantial analytic complications
nevertheless arise from ion reflections. We shall address these in a
subsequent section, but initially it is simpler to exclude those
complications by assuming the distribution to be reflectionally
symmetric in ion velocity $v$.

\subsection{Single-Humped Distributions: Density in Equilibrium}

Since $f_\infty(v_\infty)$ is symmetric in the sign of $v_\infty$,
using $vdv=v_\infty dv_\infty$ one can simply write the (ion) density
as
\begin{equation}
\label{symden}
  n = 2\int_{|v_\phi|}^\infty f_\infty {v_\infty dv_\infty\over
    \sqrt{v_\infty^2-v_\phi^2} }  
  = 2\int_{|v_\phi|}^\infty f_\infty'\sqrt{v_\infty^2-v_\phi^2}\,
  dv_\infty,
\end{equation}
where $|v_\phi|=\sqrt{2(\phi-\phi_\infty)}$ is the speed at infinity
of ions that are reflected at the position $x$, potential $\phi$; and
prime denotes differentiation with respect to argument
($v_\infty$). This density is then simply a function of $\phi$.
Express the density far from the potential structure as
\begin{equation}
n_\infty=2\int_0^\infty f_\infty dv_\infty=-2\int_0^\infty f_\infty'
v_\infty dv_\infty,
\end{equation}
and so deduce the density change introduced by the presence of the
potential structure:
\begin{equation}
  n-n_\infty=2\int_0^{|v_\phi|}f_\infty'v_\infty dv_\infty +
  2\int_{|v_\phi|}^{\infty}f_\infty'\left[v_\infty-\sqrt{v_\infty^2-v_\phi^2}\,\right] dv_\infty.
\end{equation}
When $f_\infty$ has only a single maximum (at $v=0$) $f_\infty'$ is
negative throughout the integrals. The functions multiplying
$f_\infty'$ in the integrands are everywhere positive; so for
symmetric single-humped $f_\infty$, we have $n<n_\infty$: the ion
density change arising from a positive potential is always
negative. This is one indication that a positive potential soliton
sustained by ions cannot exist at low speed relative to the ion
thermal (or acoustic) speed. The density perturbation has the wrong
polarity for self sustainment. What is more, as we shall see in the
next section, this observation has important consequences for the
possibility of slow electron holes.  For single-humped $f_\infty$ they
will have negative ion charge relative to the external plasma. This
negative ion charge repells the electron hole that causes them. The result is rapid
acceleration of the electron hole until it has speed higher than
typical ion thermal speeds. Such unstable acceleration has been well
documented in simulations\cite{Eliasson2004,Zhou2016}\footnote{A
  recent paper\protect{\cite{Mandal2020}} reports Vlasov simulations
  appearing to show for Maxwellian distributions at $T_i/T_e=10$ that
  self-acceleration is suppressed. It claims that high enough ion
  temperature $T_i/T_e>3.5$ can reverse the ion density response. That
  claim is proven by the present simple derivation to be incorrect.
  The simulation code is not initialized self consistently, resulting
  in potential oscillations much larger than the extremely small hole
  potential; these factors cast doubt on its results, which contradict
  several other code simulations at larger amplitude. The analysis
  included to explain its results is faulty.}.  Thus the slow electron
hole equilibrium in a single-humped ion distribution is unstable to
hole acceleration.

\subsection{Force and acceleration of the potential structure}

To make a more quantitative assessment of slow hole dynamics, it is
simplest to find the total force exerted on the ions by the entire
potential profile $\phi(x)$ (per unit area perpendicular to
$x$). Evidently it is
$F=\int \rho Edx=-\int_{-\infty}^{\infty} n(x) {d\phi\over dx}dx
=-\int n(\phi) d\phi$. When $f_\infty$ is symmetric, $n(\phi)$ is
independent of the sign of $x$, denoted $\sigma_x$, while $d\phi/dx$ has sign
$-\sigma_x$. Therefore, in steady state regardless of the shape of
$\phi(x)$, the total force on the ions is zero, as a consequence of
symmetry. Perhaps more significantly, the reaction force exerted by
the ions on the potential structure ($-F$) is also zero. It is in
equilibrium.

Suppose, however, that the potential structure is stationary
$\phi_0(x)$ (in the equilibrium frame) and remains in steady
equilibrium long enough for the ion density to reach the value given
by eq.\ (\ref{symden}); but then some perturbative uniform
displacement $\delta x$ of the potential structure (in the equilibrium
frame) begins, which is rapid relative to the timescale of adjustment
of the ion density to the movement. This presumption is a good
approximation for an electron hole experiencing unstable acceleration
as has been shown analytically\cite{Hutchinson2016}, and by
simulation\cite{Zhou2016} elsewhere.  Also, the timescale for electron
motion and hence structure motion is much shorter (by
$\sim \sqrt{m_e/m_i}$) than for ion motion. After a short time, the
density of the slowly responding ions, $n(x)$, will to lowest order be
unchanged, it remains a function of the steady potential $\phi_0$, but
will no longer be a function of the instantaneous potential
$\phi=\phi_0+\delta\phi$. Consequently the symmetry is broken, and
total force on the ions will be non-zero.  The linearized perturbation
for a small rigid shift $\delta x$ of the potential
structure\footnote{A pure shift of the potential structure is
  justified if the ion charge contribution to the equilibrium hole is
  small compared with the electron. If not, then the magnitude of the
  force increment will be only approximate; nevertheless, its sign,
  which determines stability, will not be changed.} is
$\delta\phi\simeq-{d\phi_0\over dx}\delta x$ which is
anti-symmetric. The ion force increment is
\begin{equation}
  \delta F = -\int n(x) {d\delta\phi\over dx} dx 
  =\delta x \int n(x) {d^2\phi_0\over dx^2} dx 
  =-\delta x \int {dn\over dx} {d\phi_0\over dx} dx
  =-\delta x \int{dn\over d\phi_0}\left(d\phi_0 \over dx\right)^2 dx.
\end{equation}
Now the potential structure has been displaced from equilibrium and
experiences a force $-\delta F= -C \delta x$, where the coefficient is
$C=\delta F/\delta x=-\int{dn\over d\phi_0}\left(d\phi_0 \over
  dx\right)^2 dx$. Incidentally, this force is related to imbalanced
reflection of the ions from the potential structure and ion
jetting. Imbalanced reflection of electrons from small
negative-potential \emph{ion holes} is proportional to the slope of the
electron distribution function at the hole speed\cite{Dupree1983}. But
for positive structures it is better to express the force in terms of
these instantaneous integrals, since full reflection of ions takes
much longer to transfer their momentum to the structure than the
timescale of hole motion. Whether or not the structure regarded as a
rigid composite object continues to be displaced or returns to its
equilibrium position depends upon the sign of its acceleration, and
hence on the sign of $C$, which is evidently minus the sign of
$dn/d\phi_0$ (averaged over the hole with positive definite weight);
but it also depends on the structure's response to force, that is, its
effective mass $M$.

Supposing the potential structure to be an electron hole, one can
deduce the acceleration of the hole by requiring the total of electron
($\dot P_e$) and ion ($\dot P_i$) momentum rates of change to be zero
(since the electric field momentum is negligible):
$0=\dot P_i+\dot P_e=\delta F +\dot P_e$. Thus the effective mass of
the hole, its force divided by acceleration, is
$M=-\delta F/\ddot{\delta x}=\dot P_e/\ddot{\delta x}$. The electron
momentum change arises from jetting by the accelerating potential
structure, and is given by equation (34) of reference\cite{Hutchinson2016}
in dimensional units

\begin{equation}
  \dot P_e = - \ddot{\delta x}\;n_e m_e\int h(\chi) dx ,
\end{equation}
where $h(\chi)$ is a non-negative function\footnote{Strictly, eq.\
  (\ref{hfunc}) applies when ion charge response is neglected. That
  neglect is not immediately obvious for slow holes. However, at the
  threshold of instability the ion charge response actually \emph{is}
  negligible, so it is appropriate to invoke the equation for
  thresholds. When estimates of unstable positive growth rate
  $\gamma\not=0$ are later obtained from it, one should beware of the
  approximation and take these as approximate only.} of argument
$\chi\equiv\sqrt{|e\phi|/T_e}$. For negligibly shifted Maxwellian
electrons $h$ can be written in closed form as
\begin{equation}\label{hfunc}
  h(\chi)=-{2\over \pi}\chi +\left[\left(2\chi^2-1\right){\rm e}^{\chi^2}{\rm
      erfc}(\chi)+1  \right].
\end{equation}
The effective electron hole mass (per unit transverse area)
$M=- n_em_e \int h(\chi) dx$ is thus negative.

Within the present lumped approximation, the equation of motion of the
potential structure is $\ddot{\delta x}=-(C/M)\delta x$, giving
eigenfrequency $\omega =\pm\sqrt{C/M}$. The \emph{stability} of the
initial symmetric equilibrium depends on the sign of $C/M$.  Stable
oscillation is expected for $C/M$ positive, exponential growth for
$C/M$ negative.  When ion density is decreased by positive
potential, $dn/d\phi_0<0$, $C$ is positive.  Therefore $C/M$ is negative
and the hole is unstable to displacements relative to the equilibrium
position and velocity when ion density change caused by positive
potential is negative, as it is for symmetric single-humped ion
distribution function.

The mass of a hole of small $\psi$ can be calculated using the
approximation $h(\chi)\to \chi^2=\phi T_0/T_e$ so in dimensional units
$M=-(n_em_e/T_e)\int e\phi dx$, and
\begin{equation}
  \label{eq:CbyM}
  {C\over M}= {-e\int {dn\over d\phi_0}\left(d\phi_0\over dx\right)^2 dx
    \over -(n_em_ee/T_e)\int \phi dx }.
\end{equation}
Now we convert $\phi,\ \psi$, $x$, and $C/M$ into dimensionless units
dividing them by $T_0/e$, $\lambda_D$, and $\omega_{pi}^2$
respectively, and defining an effective dimensionless hole length
$L\equiv\int\phi dx/\psi$. This yields the dimensionless form
\begin{equation}
  \label{eq:CbyMless}
   {C\over M}= {-e\int {dn\over d\phi_0}\left(d\phi_0\over dx\right)^2 dx
     \over -(n_em_ee/T_e)\int \phi dx } \times {1\over
     \lambda_D^2 \omega_{pi}^2}
   \simeq -2 {T_e\over T_0} {m_i\over m_e} {\psi \over L^2},
 \end{equation}
 where, anticipating a result to be shown in section \ref{section2.3},
 the dimensionless magnitude of the numerator for hole form
 $\psi\, {\rm sech}^4(x/\ell)$ and unit density Maxwellian ions is
 found to be approximately $C=2\psi^2/L$.  A self-consistent
 small-amplitude electron hole of the sech$^4$ form has length
 $L=(16/3)\lambda_{De}= (16/3)\sqrt{T_e/T_0}$ (dimensionless).
 Therefore $C/M=-(3/16)^22\psi (m_i/m_e)$ and the growth rate is
 $\gamma=\sqrt{2\psi m_i/m_e}(3/16)$ in $\omega_{pi}$ units. It is
 perhaps more intuitive to write dimensionally
 \begin{equation}
   \label{eq:gamma}
 \gamma\simeq{3\over 16}\sqrt{2e\psi\over T_0}\,\omega_{pe}.   
 \end{equation}
 This confirms that the
 instability is fast because it is on the electron time-scale
 $\omega_{pe}^{-1}$ rather than the ion timescale, hence justifying
 the model taking stationary ions during the motion of the potential
 structure. But if $\psi$ is very small, the gap between the reduced
 $\gamma$ and the ion response time will eventually disappear, and the
 approximation become inadequate.

\subsection{Non-single-humped distribution function hole stability}
\label{section2.3}

For a stable electron hole or other positive potential
structure attributable to the plasma itself to exist, we require the
ion density perturbation that it produces to be non-negative. This is
achieved in classic ion-acoustic solitons by the relative speed of the
soliton and the ions being substantially larger than the ion thermal
speed. A classic soliton is not slow in the current
sense; and also the
ion velocity distribution is not symmetric (in the structure frame)
but consists of a single Maxwellian shifted by velocity
$v_b$. Pursuing in this section only symmetric distributions, one can
clearly make the distribution symmetric by introducing a symmetric
second ion population of shift $-v_b$.  In that case a soliton can
exist, but physically it is still not ``slow'' in the sense of the
structure velocity coinciding with the dominant part of the ion
distribution.

This two-beam soliton situation shows qualitatively how to obtain positive ion
density perturbation. Ions that are not reflected, because their
energy exceeds the peak potential, contribute positively to the ion
density perturbation because their speed $|v|$ at positive potential
is lower than at $\phi_\infty$ (conserving energy) yet their flux
($nv$) must be independent of position; so $n$ must increase to
compensate. A passing monoenergetic beam of ions has density $n(\phi)=n_\infty
v_\infty/\sqrt{v_\infty^2+2(\phi-\phi_\infty)}$, which increases
without bound near the potential $\phi=\phi_\infty+v_\infty^2/2$
needed for reflection. Therefore, if the unreflected (passing) ion population is
sufficiently dominant, the ion density change is positive. What the
previous subsection showed is that for \emph{single-humped} symmetric
distributions the passing population is never sufficiently
dominant. A sufficiently widely spaced two-beam distribution can, however,
achieve sufficient dominance. If the spacing $2|v_b|$ is reduced,
eventually that dominance will be lost and the unstable negative
ion density change [$n-n_\infty$] will reappear. The intuitive question
therefore is quantitatively how small can the beam spacing be and
still avoid instability. We already know from the previous subsection
that part of the answer is that the distribution must be
non-single-humped; in other words that it must have a local minimum. But
how deep must the minimum be?

The stability threshold is determined, on the basis of the lumped
treatment of electron-sustained structure motion, by the change of sign
of the force coefficient,
$C=\delta F/\delta x=-\int{dn\over dx}{d\phi_0\over dx} dx = -\int
{dn\over d\phi_0}\left(d\phi_0\over dx\right)^2 dx$. Instability
arises if $\delta F/\delta x$ is positive. Its value is determined by
both the distribution function (giving $n(\phi_0)$) and potential
profile $\phi_0(x)$ (giving ${d\phi_0\over dx}$). However its sign
depends only on the relative shape of $\phi_0(x)$, not on its extent,
because expanding or contracting the profile in $x$ by a uniform
(positive) scale factor $\ell$ simply divides $\delta F/\delta x$ by
$\ell$.  Thus, for example, a Gaussian potential
$\phi=\psi {\rm e}^{-x^2/\ell^2}$ will give a very slightly different
threshold than $\phi(x)=\psi\, {\rm sech}^4(x/\ell)$, but neither
threshold depends on the value of $\ell$. We choose $\ell$ conveniently so
as to make $L\equiv\int \phi dx/\psi=1$, requiring $\ell=3/4$ for the
${\rm sech}^4(x/\ell)$ shape, in the following plots. Using instead a Gaussian
with $\ell=1/\sqrt{\pi}$ gives plots that appear so similar they are
not worth including. To an excellent approximation only the overall
width $L$ and height $\psi$ of the hole control the quantitative values.

To evaluate $\delta F/\delta x$ (numerically) for a specified distribution and
potential shape, we must obtain $n(\phi)$ by integrating eq.\
(\ref{symden}) with respect to $v_\infty$ and then integrate
$\int{dn\over dx}{d\phi_0\over dx} dx$ with respect to $x$, for the
chosen shape $\phi(x)$. A code has been written to perform these
integrations for arbitrary (input) $f_\infty$ or for distributions
consisting of multiple shifted Maxwellian components where
their separation (and hence the depth of the local minimum) is scanned. 

\begin{figure}[htp]
  \includegraphics[width=0.481\hsize]{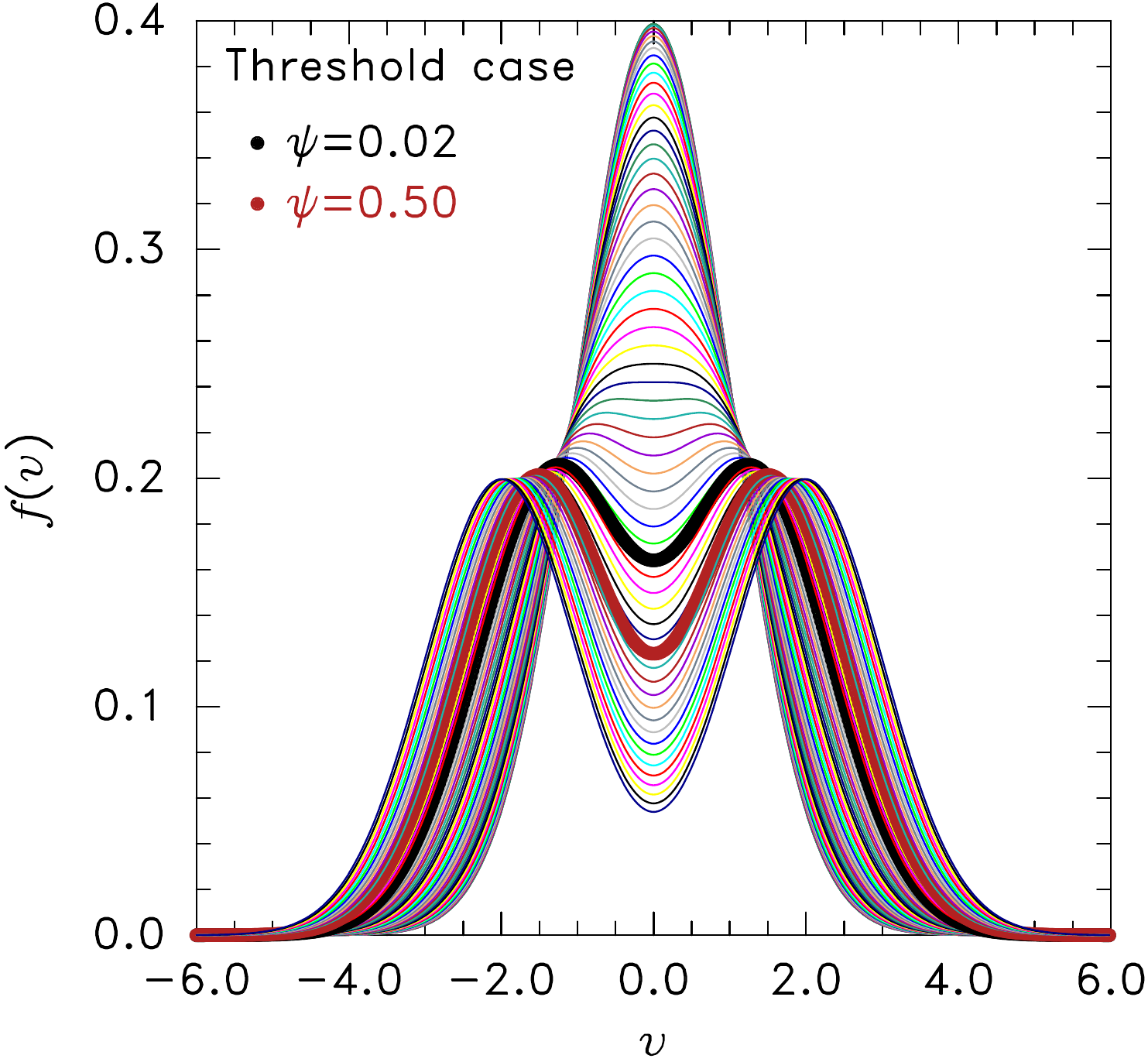}\hskip-1.5em (a)
  \includegraphics[width=0.48\hsize]{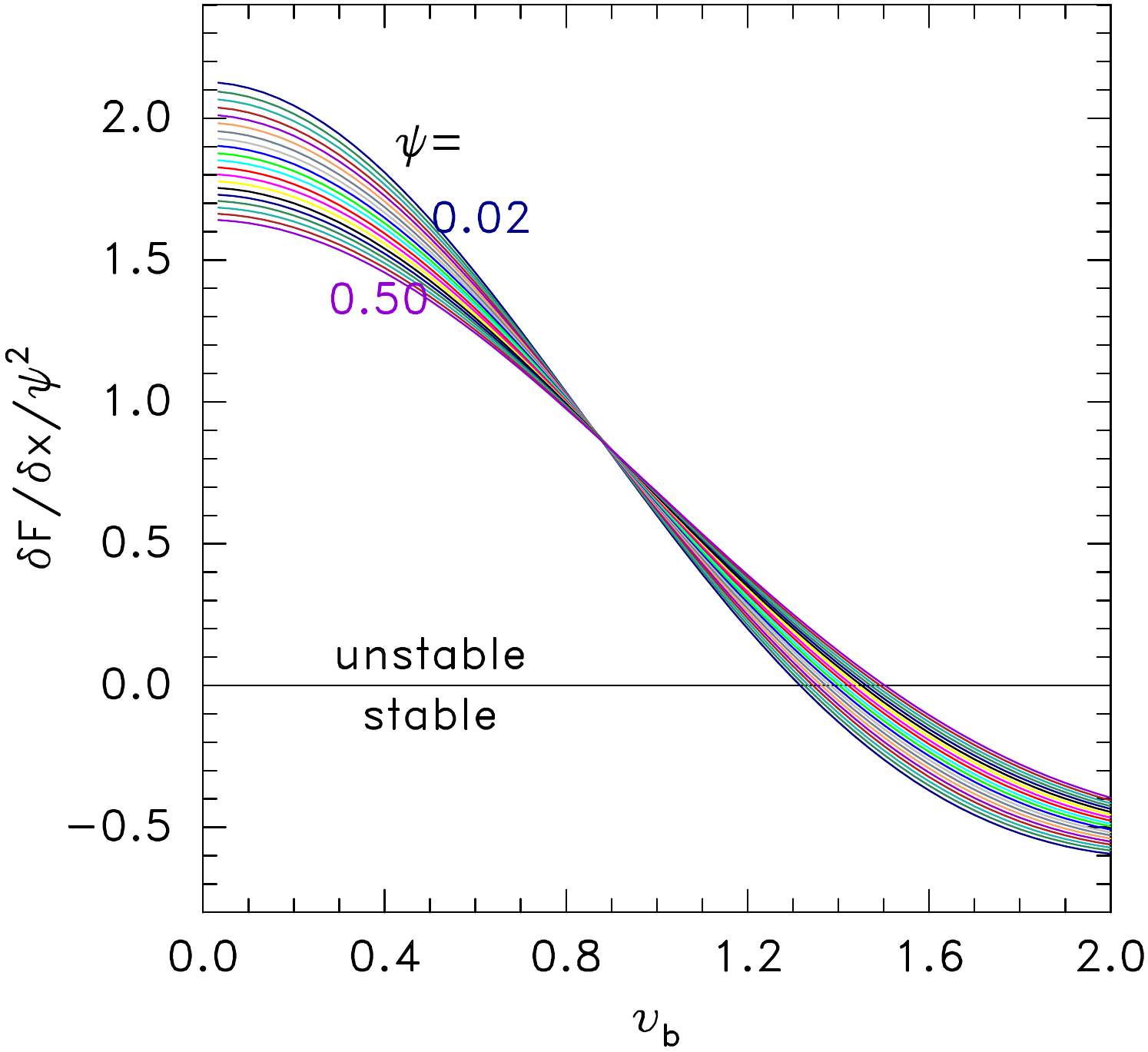}\hskip-1.5em (b)
  \caption{(a) The ion distribution functions arising for the sum of
    two symmetric Maxwellians displaced by $\pm v_b$ for
    $0<v_b<2$. The threshold case is marked in thick black for $\psi=0.5$
    and thick red for $\psi=0.02$ . (b) The force coefficient
    $\delta F/\delta x/\psi^2$ as a function of the beam
    velocity for an equally spaced range of (${\rm sech}^4$ shape) electron hole
    potential heights $\psi$.}\label{twobeamf}
\end{figure}
For distributions consisting of two symmetric Maxwellian components of
temperature $T=T_0=1$, shifted from $v=0$ by $\pm v_b$, Fig.\
\ref{twobeamf}(a) shows ion distribution shapes as $v_b$ is varied,
with the marginally stable cases for $\psi=0.02$ and $\psi=0.5$
emphasized in bold black and red. Fig.\ \ref{twobeamf}(b) shows the
force coefficients $\delta F/\delta x/\psi^2$ (using $L=1$) as a
function of beam shift $v_b$ for a range of potential heights
$\psi$. The magnitude of $\delta F/\delta x$ scales approximately like
$\psi^2$ but because of nonlinearities in $n(\phi)$ there is a small
variation in the force and threshold with potential peak height.

As shown in Fig.\ \ref{twobeamf}(b), over a
wide range of potential heights $\psi$ the threshold $v_b$, where
$\delta F/\delta x$ crosses zero, varies a modest amount: between
$1.3$ and $1.5$, and the corresponding depth of the minimum
$(f_{max}-f_{min})/f_{max}$ is a fraction of the $f$-maximum that lies
between $0.20$ and $0.36$. Greater $\psi$ requires deeper minimum.

\begin{figure}[htp]
  \centering
  \includegraphics[width=0.5\hsize]{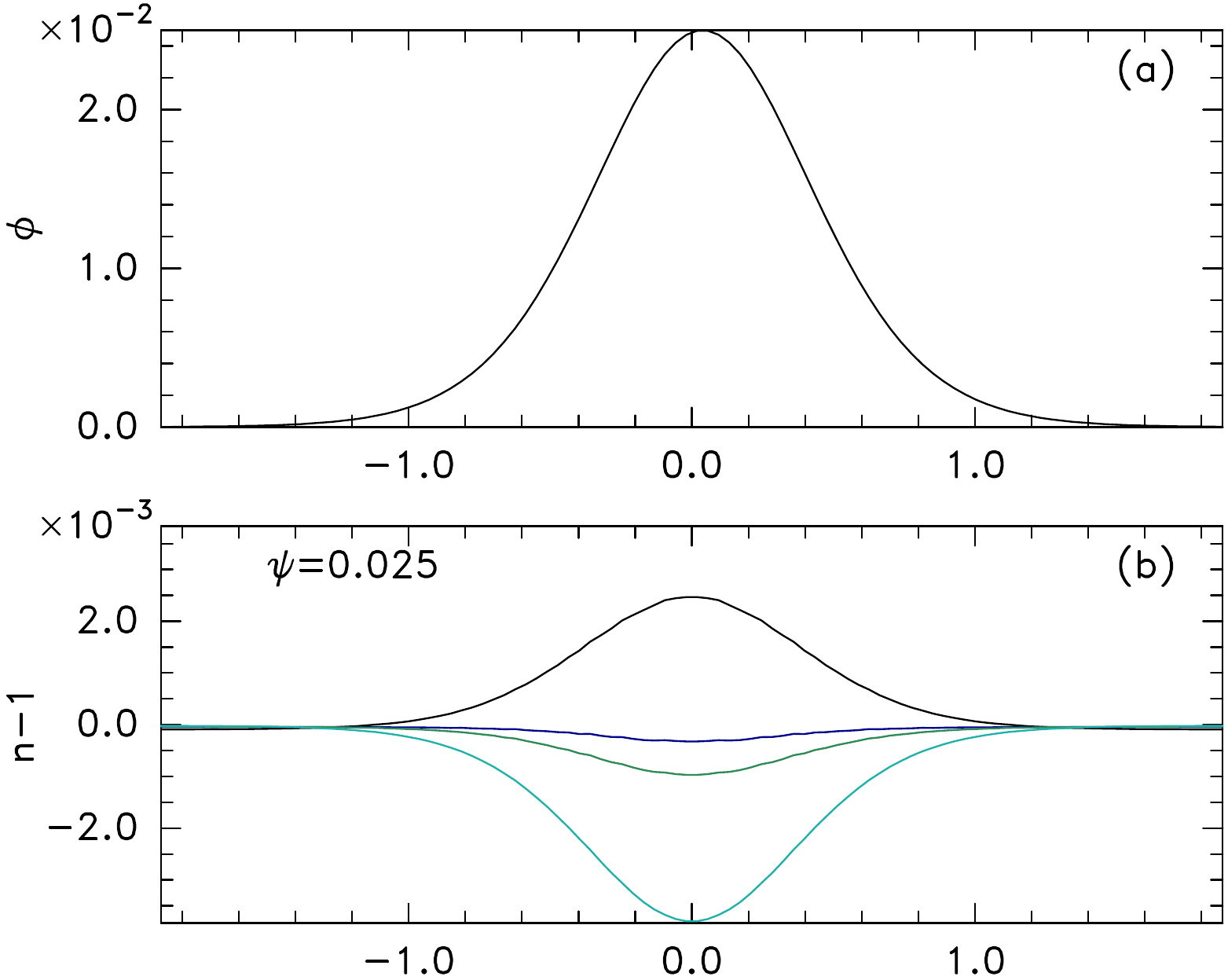}
  \includegraphics[width=0.5\hsize]{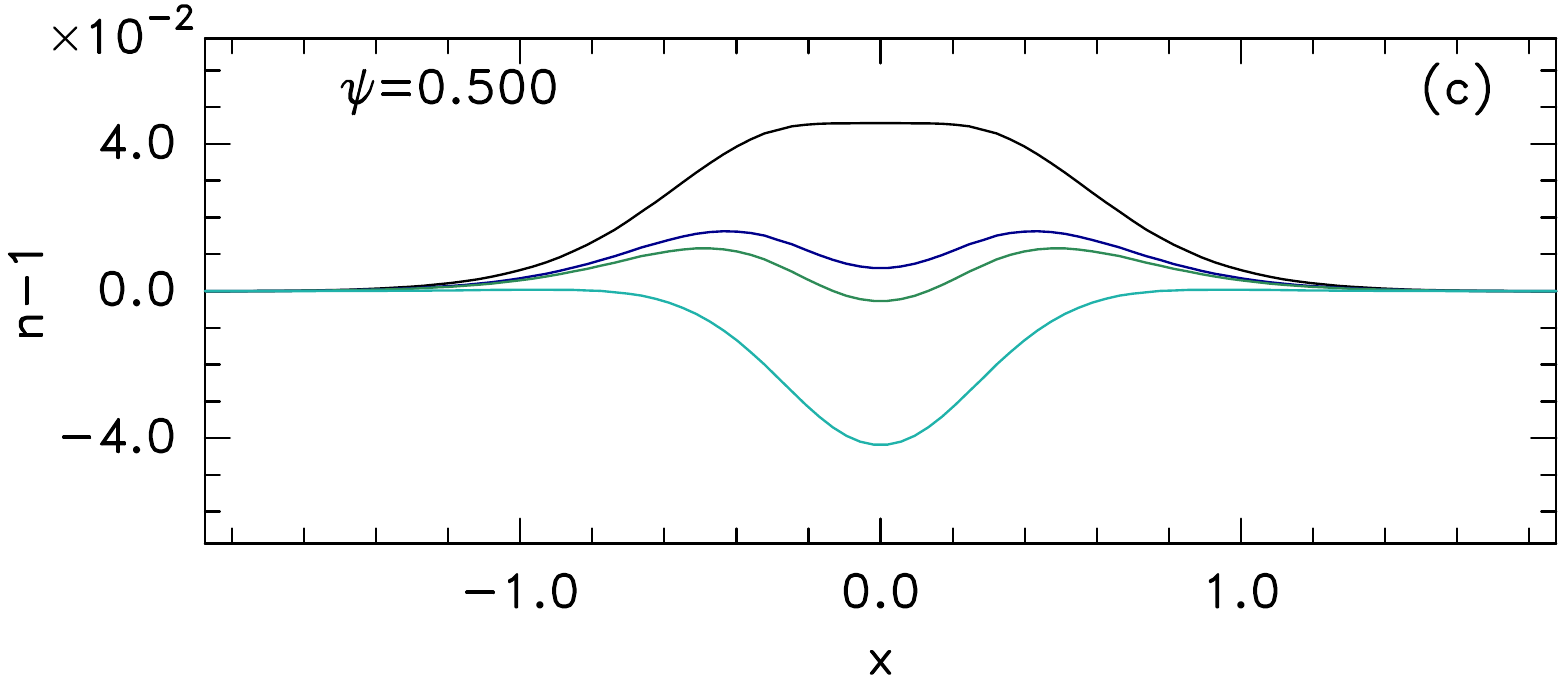}
  \caption{(a) Slightly shifted Gaussian potential profile to
    calculate $\delta F$. (b) Ion density as a function of position
    for the first stable (dark blue) and last unstable (green)
    velocity shift $v_b$, for peak potential $\psi=0.5$, together with
    cases more stable (black greater $v_b$) and more unstable (light
    blue smaller $v_b$). (c) Ion density threshold cases like (b) but
    for greater $\psi$.}\label{phinofx}
\end{figure}

Fig.\ \ref{phinofx}(a) shows for reference the used sech$^4$ potential
profile including the small shift $\delta x$ used for calculating
$\delta F$.
Fig.\ \ref{phinofx}(b) shows the (unshifted) corresponding first
stable density profile (blue) and the adjacent last unstable density
profile (green) in which the $v_b$ is smaller by $0.02$. The precise
threshold lies between these two $n(x)$ profiles, corresponding to a
density that is nearly flat (but not exactly because of profile and
nonlinear effects).  Fig.\ \ref{phinofx}(c) shows the same thing for a
much larger potential peak $\psi=0.5$. One can see that near threshold
$dn(\phi)/d\phi$ actually reverses its sign at large $\phi$.

\subsection{Positive potential structures sustained by ions?}

Since within a local $f_\infty(v)$ minimum a positive potential gives
positive ion charge, one might wonder whether such an effect can by
itself be responsible for sustaining the structure. In multiple-humped
ion distributions is there such a thing as a slow positive ion
soliton?  The answer appears to be no. The reason is not stability,
but equilibrium. To generate a solitary positive potential peak
requires the electric charge to be positive near the peak but negative
in the wings. Yes, ion charge perturbation can be positive for
positive potential whose velocity lies within a local minimum of the
distribution function; but if so it is never negative, because
actually $dn/d\phi$ decreases with increasing $\phi$. That rules out
the required transition from negative to positive ion charge
perturbation as $\phi$ rises moving from the hole wing to the
potential peak. Therefore certainly ions alone cannot sustain a slow
positive soliton. This contrasts with negative potential \emph{ion
  holes}, which can be sustained by a deficit of trapped ions.

A Maxwellian electron distribution with no trapped deficit gives a
negative charge density perturbation approximately linear with
potential ($\sim -\phi$) (see e.g.\ \cite{Hutchinson2017}).  Electrons
therefore \emph{can} give the required negative charge density in the
wings of the hypothesized solitary structure, and do so for a classic
ion-acoustic soliton. But to obtain positive charge density near the
potential peak, the rise in ion density (in a soliton) has to
overwhelm the rise in electron density in the center but not in the
wings. That requires the ion density to have substantial positive
curvature $d^2n/d\phi^2$. It has for a passing ion beam, but it
generally does not for a slow structure, because of ion reflection. In
fact (compare Fig.\ \ref{phinofx}(c)) at potentials comparable to the
width of the $f(v)$ minimum, $n(\phi)$ has substantial \emph{negative}
curvature (with $dn/d\phi$ eventually becoming negative). This
appears to be a general rule arising from the reflection of
progressively higher $f_\infty(v)$ values as the reflection velocity range
expands from a zero lying in a distribution minimum, as required for
velocity stability.

Therefore essentially any positive solitary structure that is slow in
the sense of having velocity coinciding with the dominant parts of the
ion distribution cannot be sustained by ions alone, and cannot be
sustained at all unless the electron distribution changes make major
contributions to the central positive charge. This restriction is not
exactly a watertight proof that positive slow solitary structures are
electron holes, but it closes off most plausible alternative possibilities.

\section{Asymmetric Distribution Functions}\label{sec2}

Now we must tackle asymmetric ion velocity distributions and their
complications.

\subsection{Calculation of density}

First, if $f_\infty$ is asymmetric in the incoming sign ($\sigma_{v\infty}$ say)
of $v_\infty$, the ion density will be a function of \emph{both} the
magnitude of the potential \emph{and} the side of potential peak at
which the potential occurs. That is because slow ions will be
reflected and hence contribute only on one side or the other of the
peak (positions $x$ having sign $\sigma_x=-\sigma_{v\infty}$). We should
therefore refer to the potential in a way that indicates the sign; one
convenient way to do so is to express it as the distant incoming
velocity that reflects at $\phi$:
$v_\phi\equiv-\sigma_x\sqrt{2(\phi-\phi_\infty)}=\sigma_{v\infty}\sqrt{2(\phi-\phi_\infty)}$.

But second, even with this clarification, the density is actually a
function of the local potential (and hence $v_\phi$) \emph{and also} the
height of the potential peak $\psi$; because although in a
collisionless situation $f$ and energy are constant, whether the
distribution of particles $f(x,v)$ moving away from the peak is
representative of $v_\infty$ positive or negative depends whether
those particles have been reflected or have passed over the peak.

\begin{figure}[htp]
\centering
\includegraphics[width=.8\hsize]{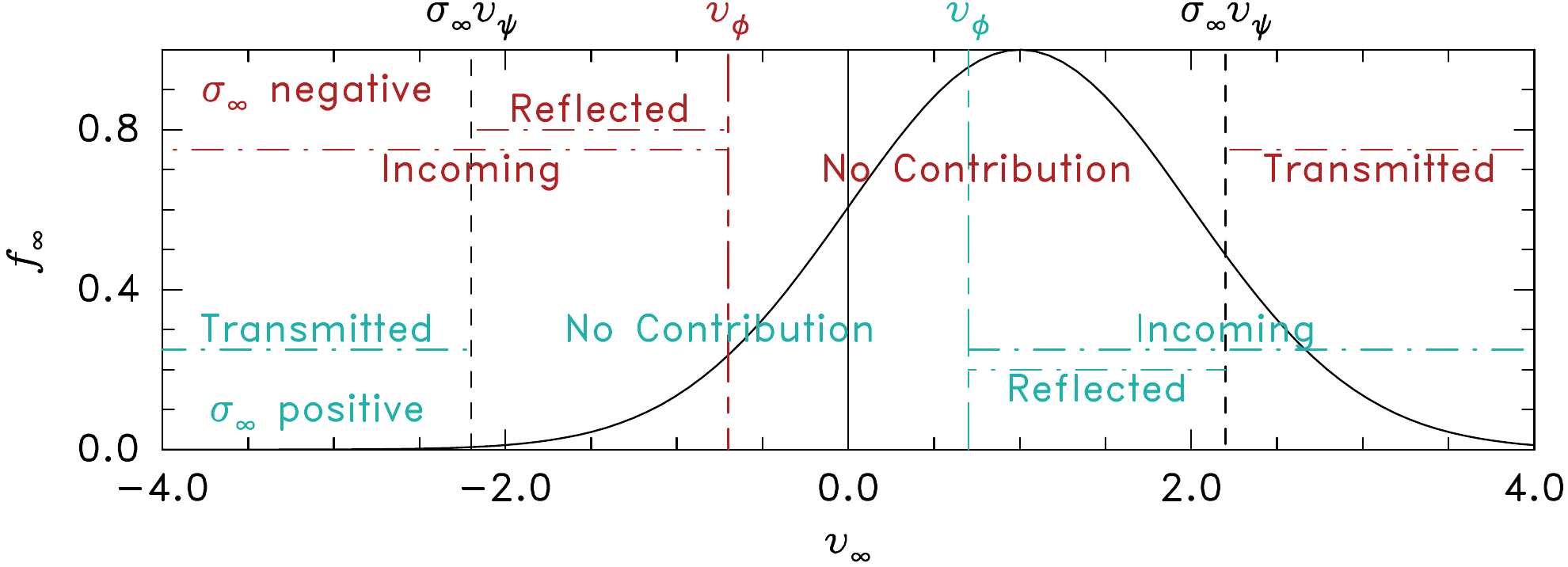}  
\caption{Distant velocity ranges ($v_\infty$, horizontal dash-dot
  lines) that contribute to the integrals of $f_\infty(v_\infty)$
  giving density, at the two sides of a potential peak: $\sigma_{v\infty}=+1$
  ($x$ negative) blue, and $\sigma_{v\infty}=-1$ ($x$ positive) red. Vertical
  lines at $v_\phi$ indicate the $v_\infty$ that reflects at potential
  $\phi$ and $v_\psi$ that reflects at $\psi$ }\label{contribs}
\end{figure}

So at potential $\phi$ whose position sign is given by $-v_\phi$, the
exiting particles ($v$ and $x$ having the same sign) have
$f_\infty(v_\infty)$ corresponding to a sign of $v_\infty$ equal to
$\mp \sigma_x$, depending on whether they have been reflected or not.
That is, for $v^2/2+\phi>\psi$ (passing particles),
$f(x,\sigma_x|v|)=
f_\infty(-\sigma_x\infty,\sigma_x\sqrt{v^2+2[\phi-\phi_\infty]}\,)$,
while for $v^2/2+\phi<\psi$ (reflected particles),
$f(x,\sigma_x|v|)=
f_\infty(\sigma_x\infty,-\sigma_x\sqrt{v^2+2[\phi-\phi_\infty]}\,)$. Since
I find this distinction requires considerable care, I illustrate it
graphically in Fig \ref{contribs}, denoting
$v_\psi=\sqrt{2(\psi-\phi_\infty)}$ (positive value).
The $v_\infty$ ranges that contribute to the
density locally at $\phi(x)=\phi$ for the two signs $\sigma_{v\infty}=+1$
($\sigma_x=-1$) blue, and $\sigma_{v\infty}=-1$ ($\sigma_x=+1$) red are
shown as horizontal dash-dot lines.
 A purely
illustrative Maxwellian $f_\infty$ is shown emphasizing that even a
shift of a symmetric Maxwellian from the structure velocity (zero)
gives rise to asymmetry, and hence dependence of density on $\psi$. 

The density is given as before by 
\begin{equation}
\int f(v) |dv|=\int f_\infty {|v_\infty|\over |v|}   |dv_\infty| 
=\int  f_\infty {|v_\infty|\over \sqrt{v_\infty^2-v_\phi^2} }  |dv_\infty|, 
\end{equation}
using $vdv=v_\infty dv_\infty$ and
$v_\phi^2\equiv 2(\phi-\phi_\infty)$, but the tricky part is the three
subranges of integration. They give density contributions we may
denote $n_r$ from particles that will be or have been reflected giving two
contributions from
velocity in $[v_\phi,\sigma_{v\infty} v_\psi]=[v_\phi,-\sigma_x v_\psi]$, $n_t$ from particles that have been
transmitted $[\sigma_xv_\psi,\sigma_x\infty]$, and $n_u$ (unreflected)
from particles that will not be reflected
$[-\sigma_xv_\psi,-\sigma_x\infty]$:
\begin{equation}
  n=n_r+n_t+n_u=\left[2\int_{v_\phi}^{-\sigma_x v_\psi}+\int_{\sigma_x v_\psi}^{\sigma_x\infty}
    +\int_{-\sigma_x v_\psi}^{-\sigma_x\infty}\right] 
  f_\infty {|v_\infty|\over \sqrt{v_\infty^2-v_\phi^2} }  |dv_\infty|.
\end{equation}
Since each integration range is increasing in absolute value, the sign
of $dv_\infty$ is the same as the sign of $v_\infty$; so
$|v_\infty||dv_\infty|= v_\infty dv_\infty$ and we need not take moduli.
For numerical evaluation it is advantageous as before (eq.\
\ref{symden}) to integrate by parts to
remove the singularity at $v_\infty^2=v_\phi^2$, giving 
\begin{equation}
\begin{split}
  n=&[f_\infty(-\sigma_xv_\psi)-f_\infty(\sigma_xv_\psi)]\sqrt{v_\psi^2-v_\phi^2}\\
  &-\left[2\int_{v_\phi}^{-\sigma_x v_\psi}+\int_{\sigma_x v_\psi}^{\sigma_x\infty}
    +\int_{-\sigma_x v_\psi}^{-\sigma_x\infty}\right]
  f_\infty'\sqrt{v_\infty^2-v_\phi^2}\, dv_\infty,
\end{split}
\end{equation}
in which the integrated parts no longer cancel.

This expression allows us to calculate the density arising everywhere
on a potential structure stationary in some inertial frame and the
resulting force on the ions
$F=-\int_{-\infty}^{\infty} n(x) {d\phi\over dx}dx= \sum_{\sigma_x=\pm1} 
\int_{\phi_\infty}^\psi n(\phi,\sigma_x) \sigma_x d\phi$. However, unlike
the symmetric case, no symmetry now tells us what the equilibrium
velocity of that frame relative to the ion distribution should be. And
for an arbitrary frame velocity $v_h$ relative to the ion
distribution, there will generally be a nonzero ion force
$F(v_h)$. Then the configuration will not be in equilibrium because
the structure potential will be subject to a net force $-F$. Only for
the particular structure velocity that makes $F(v_h)=0$ will there be
an equilibrium.  (This is true also for a distribution like a sum of
two similar but shifted Maxwellians, that has velocity symmetry in
\emph{some other} frame of reference; but we previously tacitly
adopted that particular frame of reference as our equilibrium
structure frame.)

\subsection{Velocity equilibrium and stability}
\label{3.2}

For $f_\infty$ intrinsically asymmetric we need to find both the
equilibrium electron hole velocity (potential structure velocity),
$v_{h0}$, and the derivative of the force with respect to a shift of
the structure relative to its equilibrium position,
$\delta F/\delta x$, to determine the equilibrium's stability to rapid
motion of the potential structure leaving behind a fixed ion
density. However before we do that, a second factor concerning
stability arises in respect of $dF/dv_h$. If we find an equilibrium
hole velocity, which has $F=0$, then how does $F$ vary when we
consider a neighboring hole \emph{velocity} (not \emph{position})?
This question governs the stability of the situation for slow hole
acceleration in the opposite limit where the ion density perturbation
accelerates with the potential structure. If $F$ changes in such a
direction as to oppose the acceleration, the equilibrium is stable to
such acceleration; but if not then the equilibrium is unstable. Of
course, in reality the two types of hole motion, having stationary ion
density, or having perfectly tracking ion density, are approximate
extreme limits of a continuous response dependent on frequency. Full
frequency analysis proves to be mathematically challenging even for an
ion stream that is well separated from the hole velocity, but has been
completed showing oscillatory instability for hole speed down to a few
ion sound speeds\cite{Zhou2017}.  In the present work we content
ourselves instead with the combination of a more heuristic pair of
approximations: the extreme limits of fast and slow. This renders
stability criteria but not precise eigenvalues.

We can formulate a lumped parameter treatment by supposing that we can
combine the two different perturbations of the ion force arising from
the coefficients $\delta F/\delta x$ and $dF/dv_h=dF/d\dot x$ into a
second order system:
$M\dot v_h= M\ddot x = -(\delta F/\delta x) x - (dF/d\dot x) \dot x$,
(recalling that $F$ is the force on the particles, which is minus the
force on the hole). It has the form
\begin{equation}
  \label{eq:harmosc}
  \ddot x +{dF\over Md\dot x}\dot x+{\delta F\over M\delta x} x= \ddot x+
   b\dot x + c x=0.
\end{equation}
The solutions of this linear second-order equation are stable if and
only if both $c$ and $b$ are non-negative. In that case, it is a damped
harmonic oscillator equation. When instead $c$ is negative, then an
exponentially growing solution dominates the long-time behavior. If
$c$ is positive but $b$ is negative, a growing oscillation is the
instability. Stability of the electron hole requires both $c=C/M={\delta
  F\over M\delta x}$ and $b={dF\over Md v_h}$ to be positive. And
since $M$ is negative that means the two ion force derivatives must be
negative.

Finding a stable equilibrium is carried out as follows.  For a given
distribution, the structure velocity $v_h$ is scanned in small steps
relative to the ion distribution to find the first (and usually the
only) value at which the force $F$ changes from positive to negative
($dF/dv_h<0$, making $b$ positive) \emph{and also} the value of
$\delta F/\delta x$ is negative so $c$ is positive (i.e.\ stable). If
no such $v_h$ is found, the distribution does not permit stable slow
electron holes. 

The other major approximation we make here is that we do not calculate
self-consistently the form of the electron hole potential
$\phi(x)$. For symmetric distributions, in principle we could choose
it to be whatever we like and find the required self-consistent
trapped electron distribution through the integral equation analysis
of Bernstein, Greene, and
Kruskal\cite{Bernstein1957,Hutchinson2017}. Since here we are
calculating the effects on a known potential of the interaction with
the ions, it is sufficient just to prescribe the potential, especially
since as noted in section \ref{section2.3} the detailed shape of the
potential has only a rather weak effect. However, for asymmetric ion
distributions and finite hole peak potential $\psi$, it is no longer
the case that the ion density is the same on the two sides of the
electron hole. Therefore it is far from obvious that the potential
need be the same on the two sides either. When it is not and
$\phi_\infty$ is different for $\pm\sigma_x$, the potential structure
has the form of a non-monotonic double layer\cite{Raadu1989} and many
additional complexities arise, which it is not the purpose to address
here. Therefore we continue by setting aside these complications,
assuming a symmetric potential form $\phi(x)\propto{\rm sech}^4x$ and
limiting the applicability of the present result to electron holes or
other structures that have negligible net potential drop across them.
Detailed analysis, in preparation for a future publication, shows that
this is a good approximation.

\begin{figure}[htp]
  \centering
  \includegraphics[width=.48\hsize]{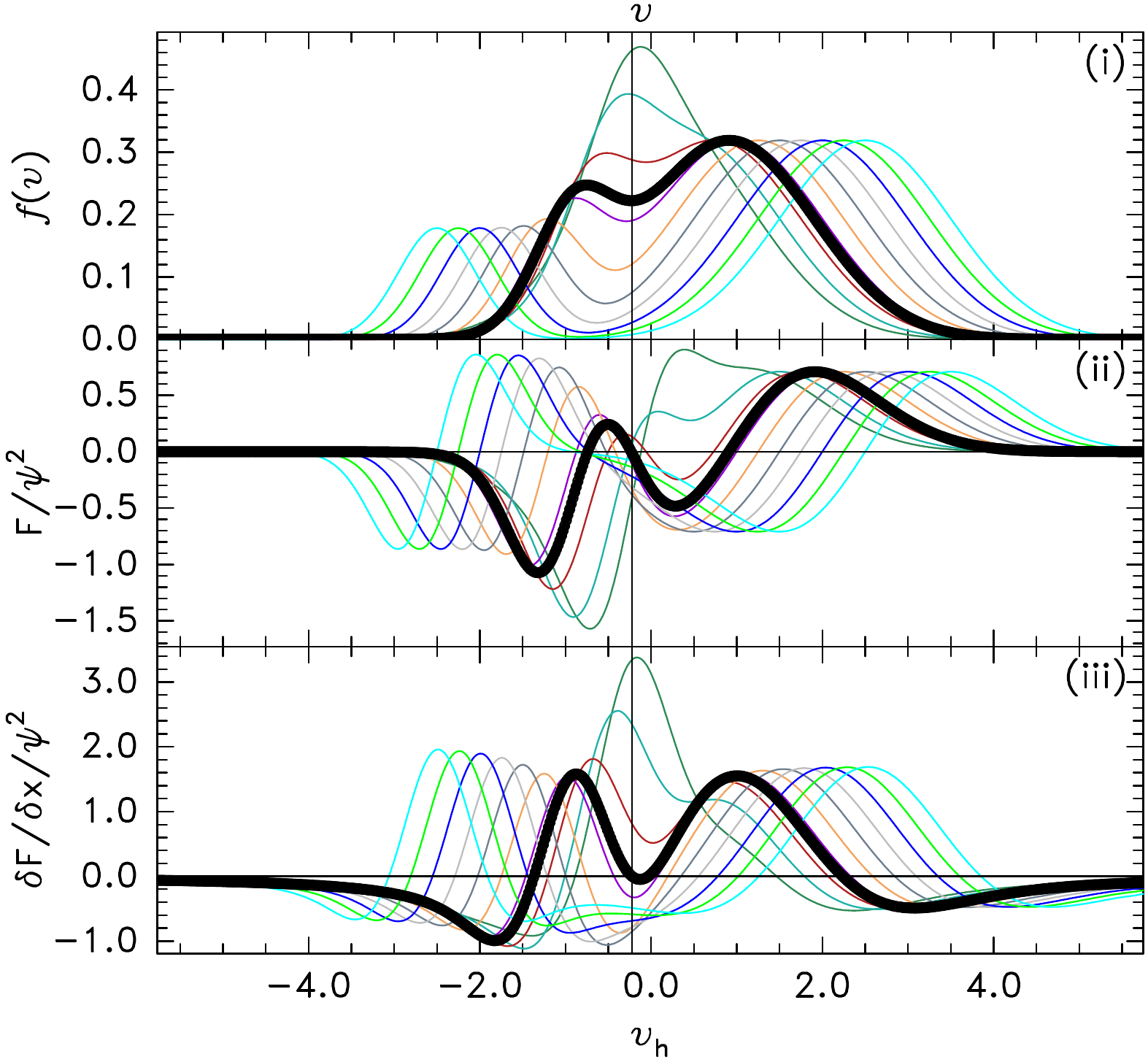}\hskip-2em(a)\hskip1em
  \includegraphics[width=.48\hsize]{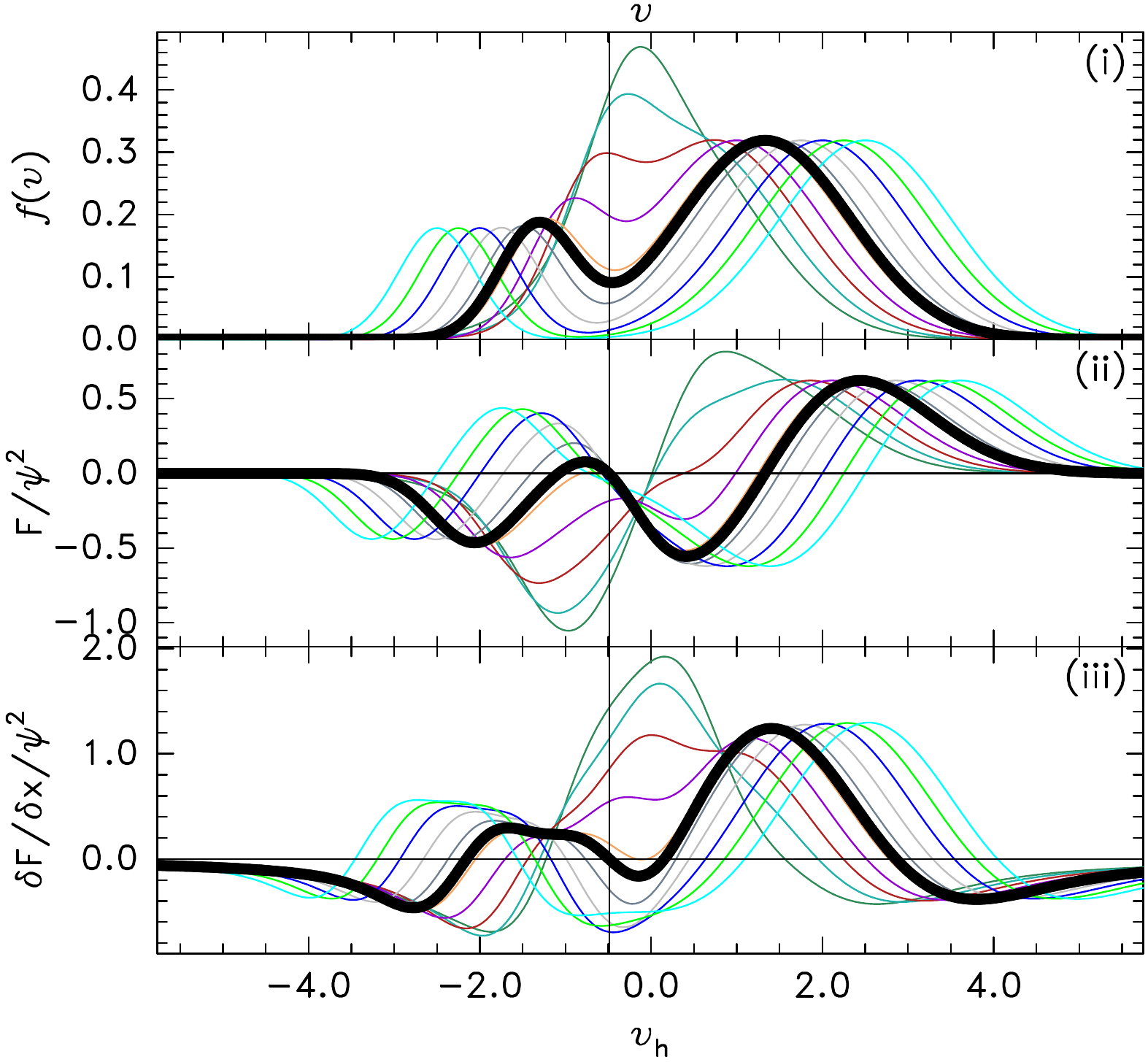}\hskip-2em(b)\hskip1em
  \caption{Distribution function (i) $f(v)$, force (ii) $F(v_h)$, and
    stability coefficient (iii) $\delta F/\delta x$, as a function of
    velocity ($v$ and $v_h$), for a range (different colors) of
    distribution beam shift parameter $v_b$. Thick black line is the
    threshold case for stable electron hole existence. (a)
    $\psi=0.02$, (b) $\psi=0.5$.}
  \label{fig:vsthreshold}
\end{figure}

The specific distribution shapes considered here are adequately
represented by the sum of two Maxwellian components shifted from each
other by $2v_b$. The widths and relative densities of the two
components can be prescribed so as to represent different generic
shapes. The shift parameter $v_b$ determines how deep any local
minimum in the distribution is. The threshold value of $v_b$,
at which stable electron holes become permitted, is found by the following
outer iteration of the above described structure velocity $v_h$ scan. A
relatively coarse $v_b$ scan is carried out over a range sufficient to
cover all desired distributions, and the smallest $v_b$ that permits a
stable equilibrium is found. An example scan is shown by the different
colored lines in Fig. \ref{fig:vsthreshold}.
The threshold $v_b$ is then refined by setting the last $v_b$ that did
not permit an equilibrium and the first that did permit it as the
lower and upper limits of a new scan of $v_b$ with the same number of
steps (hence much smaller steps). Refined scans are not plotted. This
process of decreasing the step size by a factor equal to the number of
steps in the scan is iterated; when 10 steps are taken, two more
iterations are enough to converge within other uncertainties. The
stable case of the final $v_b$ scan is interpolated for the threshold,
and is plotted in thick black. The vertical line indicates the stable
structure velocity $v_{h0}$.

It is observed in this and all other cases explored that the stable
$v_{h0}$ lies within a distribution function local depression but not
necessarily exactly at the velocity of minimum $f(v)$.
A stable electron hole equilibrium is found only if there exist three
stationary points (two maxima and one minimum between them) of $f(v)$,
and $v_{h0}$ always lies between the locations of the maxima. It is
also found that the required fractional depth of the minimum increases
as $\psi$ is increased, as the comparison between (a) and (b) of Fig.\
\ref{fig:vsthreshold} illustrates.

\begin{figure}[htp]
  \centering
  \includegraphics[width=.8\hsize]{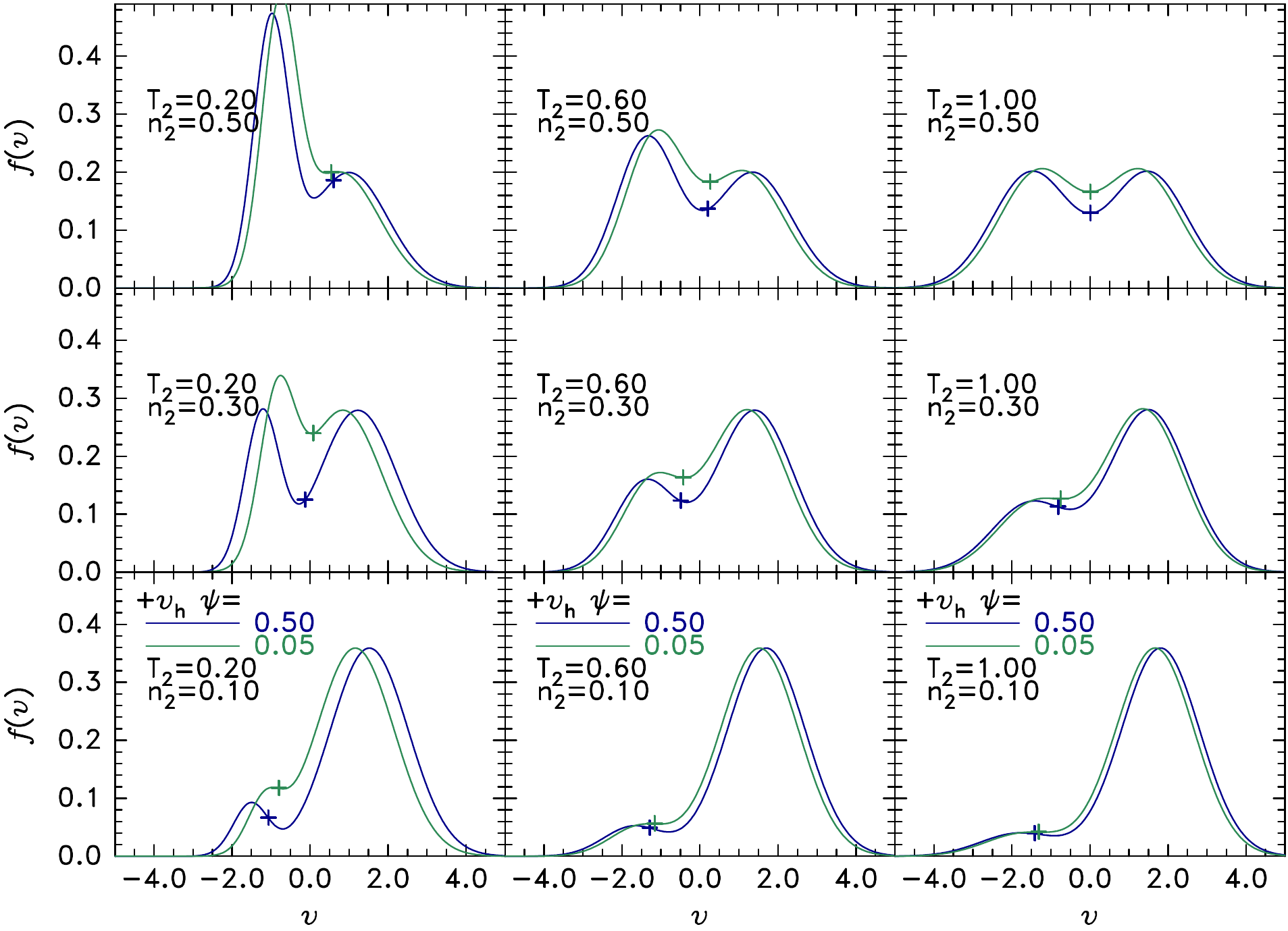}
  \caption{A range of marginal ion distribution shapes for two-Maxwellian
    distributions.}
  \label{fig:shapes}
\end{figure}

A wider-ranging graphical impression of the range of marginal
distribution shapes is given by Fig.\ \ref{fig:shapes}.
Each frame shows marginally stable distributions for $\psi=0.05$ (green) and
$\psi=0.5$ (blue), for different temperature $T_2$ and fractional
density $n_2$ of the second Maxwellian component (toward negative
$v$). The first component has $T_1=1$ and $n_1=1-n_2$. The equilibrium hole
velocity $v_{h0}$ is shown by the cross.

An estimate of the size of the damping coefficient $b$ can be obtained
by the observation that the typical value of $dF/dv_h$ in plots like
Fig.\ \ref{fig:vsthreshold} is of order $-0.5\psi^2$ or smaller. Since
the dimensionless hole mass is
$M=-(m_eT_0/m_iT_e)\int \phi dx=-(m_eT_0/m_iT_e)L\psi$ and typically
$L=(16/3)\sqrt{T_e/T_0}$ we have
\begin{equation}
  \label{eq:bcoef}
  |b|=\left|dF\over M dv_h\right| \lesssim 0.1{m_i\over m_e}\sqrt{T_e\over T_0}\psi.
\end{equation}
This is rather comparable to the maximum absolute value of $c$ in ion
dimensionless units,
\begin{equation}
  \label{eq:ccoef}
  |c|=\left|\delta F\over M \delta x\right| \lesssim \left(3\over
    16\right)^2{m_i\over m_e} \psi,
\end{equation}
for $\sqrt{T_e/T_0}$ not much different from unity. The large values
of both would mean that harmonic solutions would in fact be more than
critically damped. However, since the resulting timescales are very
short compared with $1/\omega_{pi}$, the mechanism of the ion density
perturbation perfectly changing with hole motion as if in
steady-state, giving rise to ($b$), is liable to be a very poor
quantitative approximation, unlike the stationary ion mechanism ($c$)
which is the opposite extreme. Nevertheless, the fact of $b$'s sign
being that of damping is an important indication that slow hole
motions with $c$ near zero will in fact be stable. A proper treatment
for arbitrary frequency, of course, requires solution of the
time-dependent Vlasov equation for ions, which is not attempted here.

\section{Ion Distribution Linear Stability}\label{sec3}

\begin{figure}[htp]
  \centering
  \includegraphics[width=.7\hsize]{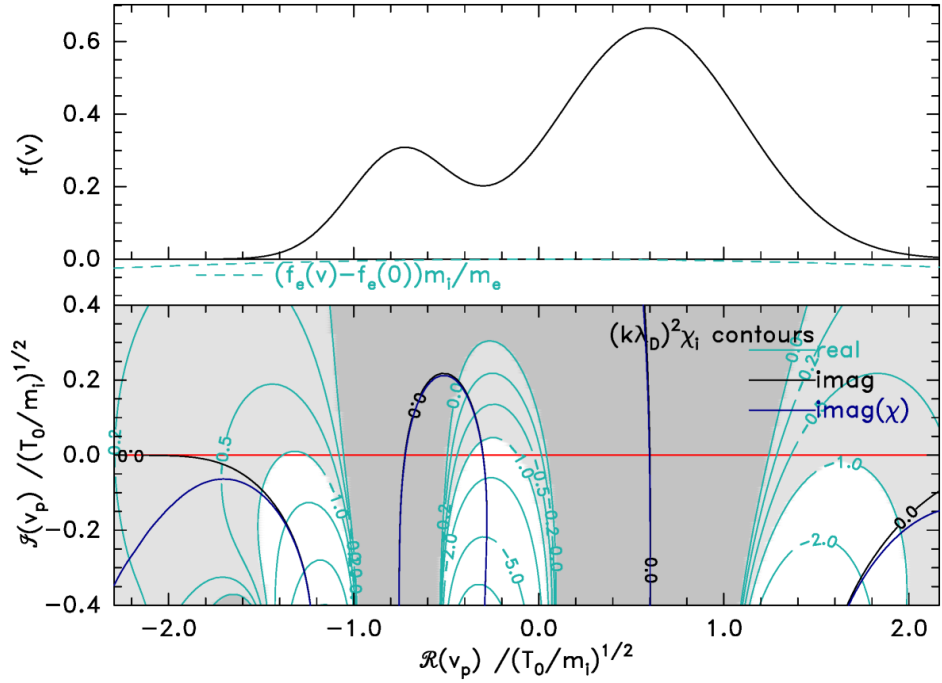}
  \caption{Contours of scaled susceptibility on the complex phase
    velocity plane. Dispersion relation solutions lie at
    intersections of the black or blue contours with the green
    contours. No solutions exist in dark shaded regions. Solutions
    exist in light shaded regions only for $T_e\ge T_0$.}
  \label{fig:ionstability}
\end{figure}

The background one-dimensional warm two-beam ion distributions might
experience sinusoidal linear electrostatic instability not caused by
solitary structures, depending on the spacing, relative density, and
velocity-width of the beams, and on the electron
temperature\cite{Stringer1964,Fried1966}. Generally this ion-ion
instability requires a local minimum in the ion distribution of a
certain depth. It is conceivable that such an instability might
contribute to the mechanism that \emph{forms} an electron hole, but
that is not the focus here. However, the question arises as to whether
the non-single-humped background plasma ion distribution
\emph{required for persistence} of slow electron holes is stable to
ion-ion modes. If not, perhaps electron hole existence would be
prevented because the required background distribution minimum is
unstable. The ion-ion instability can be explained in outline, with reference
to Fig \ref{fig:ionstability}, obtained by a numerical method of
analyzing and visualizing these and other kinetic ion instabilities
for arbitrary ion distributions developed by the present author,
mostly for pedagogic purposes [see
\url{https://github.com/ihutch/chiofv}].
The one-dimensional ion velocity distribution being analysed is shown
in the upper panel. The small middle panel shows the peak of the
electron distribution function with temperature $T_e=T_0$, to indicate
its very different velocity scale and the limited extent to which its gradient
is significant. 

For a wave of complex frequency $\omega$, real wave number $k$, and
thus phase velocity $v_{p}=\omega/k$ the lower panel shows contours on
the complex phase velocity plane (scaled to an ion speed
$\sqrt{T_0/m_i}$ where $T_0$ is a reference temperature) of the
complex quantity $(k\lambda_D)^2\chi_i$; the Debye length is
$\lambda_D=\sqrt{\epsilon_0T_0/e^2n_e}$, and $\chi_i$ is the ion
susceptibility. For species $j$, the quantity
\begin{equation}\label{ionint}
  k^2\lambda_{D}^2 \chi_j = - (\lambda_D\omega_{pj})^2 \int  {\partial
    f_j\over \partial v } {1\over v- v_p} dv\ ,
\end{equation}
(integrating along the Landau contour) is a function only of the ion
distribution shape and (complex) $v_{p}$, not of $\omega$ and $k$
separately. The dispersion relation for electrostatic waves is
$\chi=\chi_e+\chi_i=-1$. Therefore any solution must have the
imaginary part of $\chi$ equal to zero. Contours of zero imaginary
part of $\chi_i$ and of $\chi$ are shown in black and blue
respectively. The electron susceptibility (for Maxwellian electron
temperature $T_e$) gives to an excellent approximation
$(k\lambda_D)^2\chi_e =(T_0/T_e)(1+iv_{p}\sqrt{\pi m_e/2T_e})$, and
contributes an adjustment to the imaginary part of $\chi$ at large
$|\Re(v_{p})|$, which is electron Landau damping of ion-acoustic
waves. But for the low velocity instability this contribution is
negligible and the black and blue contours coincide.

The electron contribution to the real part of $(k\lambda_D)^2\chi$ is
to an excellent approximation simply $(T_0/T_e)$. And the real part of
the dispersion relation is then
$k^2\lambda_D^2 = -\Re(k^2\lambda_D^2\chi_i )-T_0/T_e$; so the
intersection of the zero imaginary contour with a green contour
indicates the value of dispersion solution's wavenumber. If $k$ is
regarded as a free choice, the dispersion relation can be satisfied
for some $k$ if $\Re(k^2\lambda_D^2\chi_i )+T_0/T_e$ is
negative. Therefore the limiting solution of the dispersion relation
as $k\to 0$ lies at $\Re(k^2\lambda_D^2\chi_i )=-T_0/T_e$, and when
$T_e/T_0\to \infty$ it is at $\Re(k^2\lambda_D^2\chi_i)=0$. Where the
appropriate contour of $\Re(k^2\lambda_D^2\chi_i)$ crosses the zero
contour of $\Im(\chi)$ is the dispersion solution for $v_{p}$. If the
intersection lies below the real axis, the mode is damped; if above,
it is growing (unstable). In regions where $\Re(k^2\lambda_D^2\chi_i)$
is positive, no solutions exist (regardless of non-negative electron
temperature) and the plane is shaded dark gray. In regions where
$\Re(k^2\lambda_D^2\chi_i)>-1$ no solution exists for $T_e\le T_0$,
and the regions where it lies between $-1$ and $0$ are shaded light
gray. The green contours for different negative values
$-$(0,.2,.5,1,2,5,10) of
$\Re(k^2\lambda_D^2\chi_i)$ therefore correspond to boundaries of
solution regions for $T_e=-T_0/\Re(k^2\lambda_D^2\chi_i)$.

Two-beam distributions like Fig.\ \ref{fig:ionstability} have three
unshaded regions where solutions exist. Those on the right and left
are the positively and negatively propagating ion-acoustic waves,
lying below the real axis when electron Landau damping is
included. The central region coinciding with the distribution local
minimum is where ion-ion instability solutions lie.
The stability of this ion-ion
mode is not significantly influenced by electron Landau damping. Changing
the electron temperature changes the stability not by changing the
local electron distribution \emph{gradient} but by changing its
\emph{height} (inversely with its overall width).  

There is therefore a threshold electron temperature above which a
(sufficiently) double-humped ion distribution function becomes
unstable. This temperature is a convenient way to parameterize the
ion-ion stability of the distribution, and it can be found simply by
examining $k^2\lambda_D^2\chi_i$ along the real $v_p$ axis, and
finding its real value at the velocity where its imaginary part is
zero, giving $T_{threshold}=-T_0/\Re(k^2\lambda_D^2\chi_i)$. This is
equivalent to the standard Nyquist stability analysis used in this
context by Penrose\cite{Penrose1960}, but expressed in a different way.

\begin{figure}[htp]
  \centering
  \includegraphics[width=0.48\hsize]{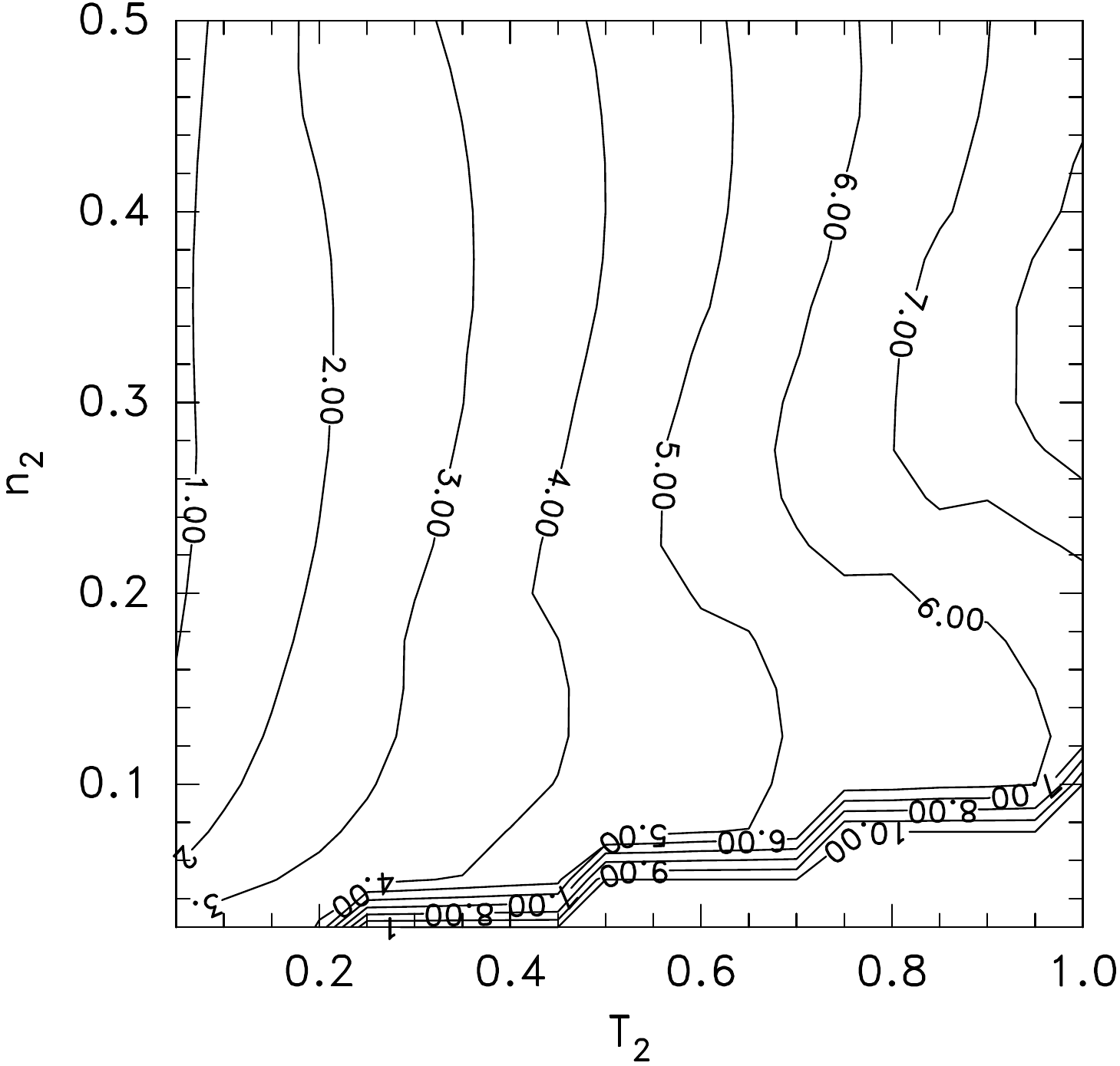}\hskip-2em(a)
  \includegraphics[width=0.48\hsize]{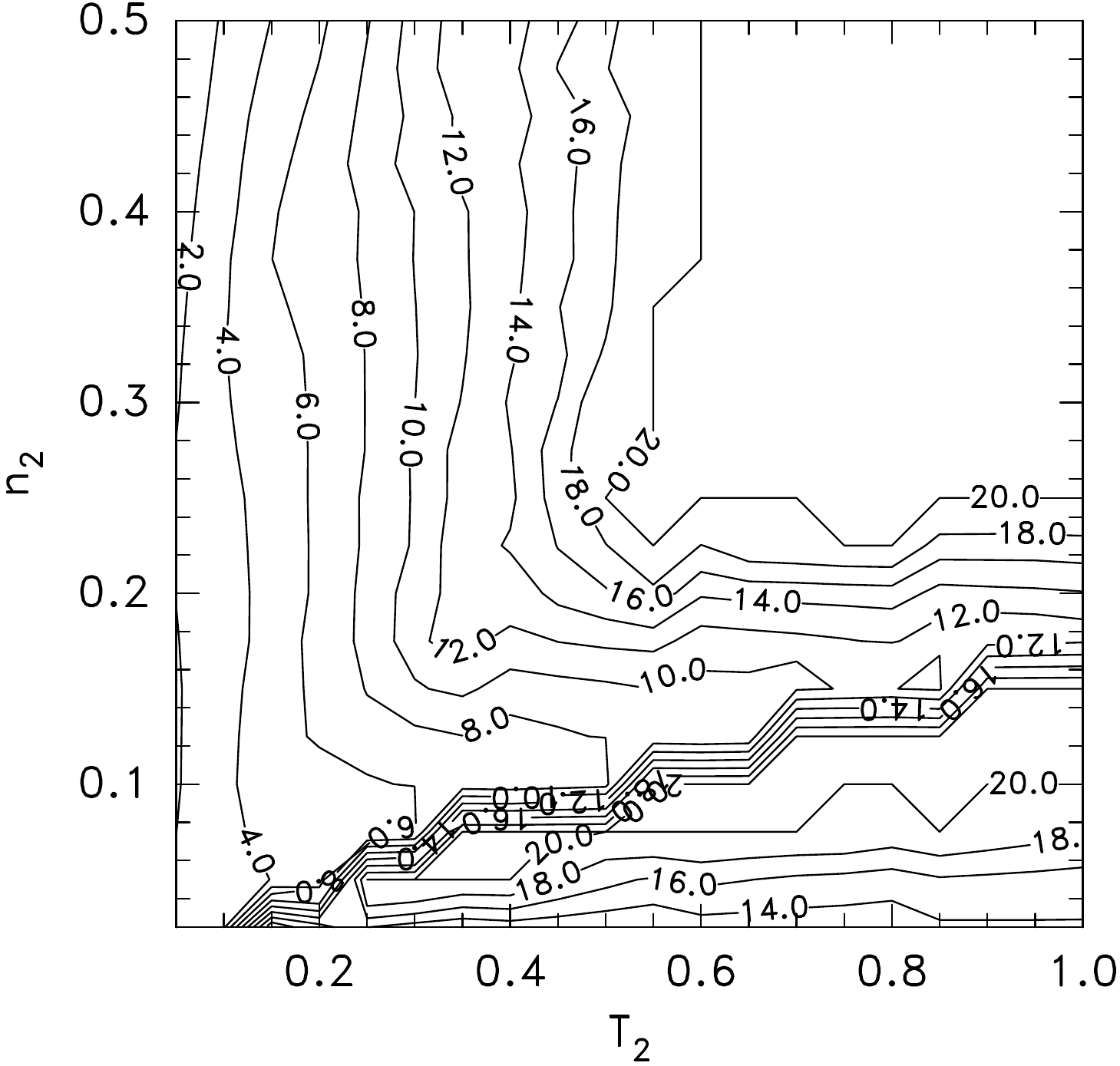}\hskip-2em(b)
  \caption{Contours of the electron temperature $T_{threshold}$ above
    which ion-ion instability occurs in two-Maxwellian
    distributions for which the first component has unit temperature
    and the second component temperature $T_2$ and fractional density
    $n_2$, and the shift between them is minimal for the existence of
    electron holes having (a) $\psi=0.5$, and (b) $\psi=0.1$.}
  \label{fig:thresh}
\end{figure}

Fig. \ref{fig:thresh} shows contours of $T_{threshold}$ as a function
of the distribution parameters. Those parameters are ordered like
Fig.\ \ref{fig:shapes}, which therefore shows qualitatively how the
distribution shape changes over the contour plane. Higher
$T_{threshold}$ are more stable cases.
The first component of the two-Maxwellian distribution has temperature
$T_0=1$, and density $1-n_2$, where $n_2$ is the total density of the
second component, whose temperature is $T_2$. Thus $n_2$ and $T_2$
together determine the shape of the distribution, because the
component velocity separation $2|v_b|$ is found by the process
described in section \ref{3.2}. In other words $|v_b|$ is taken to be
the minimum that allows the persistence of an electron hole. \emph{That} beam
separation is never large enough that the purely
one-dimensional ion-ion mode is more stable than the obliquely
propagating mode (which can happen for large
$|v_b|$\cite{Forslund1970a}); so if the ion distribution is stable by
the one-dimensional analysis, it is stable to all unmagnetized
electrostatic ion-ion modes.

Fig.\ \ref{fig:thresh}(a) is for a large amplitude hole $\psi=0.5$
which requires substantially deeper minimum in the distribution
(larger $|v_b|$ and hence less stable) than Fig.\
\ref{fig:thresh}(b). The value $\psi=0.1$ (b) is representative of
small $\psi$. The uncertainty level of perhaps a few percent in the
threshold away from the sharp cliff probably arises from the discrete
velocity meshes and from the relatively coarse parameter mesh
$20\times 20$.  Since large $T_{threshold}$ occurs for some regions,
contours are not shown above $T_{threshold}=10$ (a) or 20 (b).

At the rather large (and hence more
unstable) amplitude $\psi=0.5$ (a), even for unusually small $T_2$ no
ion-ion instability occurs until $T_e\gtrsim T_0$ or in fact
$T_e\gtrsim 6 T_{min}$, where $T_{min}$ is the smaller of the two
components' ion temperatures, no matter what $n_2$ is. For small
amplitude holes ($\psi=0.1$) (b), $T_{threshold}\gtrsim 20 T_{min}$
over the great majority of the plane, showing that the marginal
distribution for the existence of small amplitude holes is highly
stable to the ion-ion mode.

\section{Conclusions}

Long-lived one-dimensionally stable slow electron hole equilibria can
exist only when the background ion velocity distribution has a
sufficiently deep local minimum and the electron hole speed lies
within it. If the electron temperature is less than $\sim 6$ to 20
times the effective temperature of the colder ion component, then the
required background ion distribution will be linearly stable; so there
is no (linear electrostatic, 1-D) stability reason it should not
exist. The ion density change caused by a solitary positive potential
peak whose velocity lies in the local minimum, is positive, avoiding
the self-acceleration of the hole that otherwise occurs. However, ion
charge perturbations alone cannot create the conditions for a
\emph{slow} positive soliton, and the electron charge perturbation of
a distribution without phase-space-density deficit in the trapped
region cannot permit a total charge density positive at the potential
peak and negative in the wings, as is required for a
soliton. Therefore is seems that persistent, slow, positive, solitary
potential structures must be sustained primarily by trapped electron
deficit. That is, they must be electron holes. And their velocity must
lie within a local minimum in the ion velocity distribution. It is not
impossible that slow electron holes might be observed as they form, or
shortly afterwards, in ion distributions that do not possess the local
minimum found here. But they would be expected to be unstable, and so
rapidly be self-accelerated to speeds that are no longer slow.

All of the analysis presented here is purely one-dimensional. However,
it seems possible that the ion coupling effects explored might also
have a significant effect on the multidimensional transverse stability
of electron holes, by altering the force-balance that determines
it\cite{Hutchinson2019}. If so, which is a possible topic for future
analysis, they might have different typical transverse sizes than fast
electron holes, or even persist at lower magnetic field strengths.

\section*{Acknowledgements}

I am grateful to I Y Vasko, Y Kamaletdinov, and A V Artemyev for
stimulating discussions of slow electron holes, especially their
recent analysis of MMS observations confirming that they lie in minima
of $f_i(v)$. The present work was not supported by any external public
funding. The code used to calculate and plot the figures may be found
at \url{https://github.com/ihutch/slowholes}; no data was used.

\bibliography{JabRef}

\end{document}